\documentclass[prl,amsmath,amssymb,twocolumn, aps,superscriptaddress,floatfix,preprintnumbers,10pt,notitlepage]{revtex4-1}
\RequirePackage[switch,columnwise]{lineno}
\usepackage{times}
\usepackage{graphicx}
\usepackage{amsmath, braket, amsfonts}
\usepackage{amssymb}
\usepackage{natbib}
\usepackage{sidecap}
\usepackage{bm, color, ulem}
\usepackage{comment}
\usepackage{xcolor}
\usepackage{tikz}
\usepackage{pgfplots}
\usepackage{setspace}
\usepackage{titlesec}
\pgfplotsset{compat=1.12}
\titleformat*{\section}{\normalsize\bfseries}

\begin{document}
\title{Decoupling superconductivity and correlated insulators in twisted bilayer graphene}
\author{Yu Saito}
\affiliation{California NanoSystems Institute, University of California at Santa Barbara, Santa Barbara CA 93106, USA}
\author{Jingyuan Ge}
\affiliation{Department of Physics, University of California at Santa Barbara, Santa Barbara CA 93106, USA}
\author{Kenji Watanabe}
\affiliation{National Institute for Materials Science, 1-1 Namiki, Tsukuba 305-0044, Japan}
\author{Takashi Taniguchi}
\affiliation{National Institute for Materials Science, 1-1 Namiki, Tsukuba 305-0044, Japan}
\author{Andrea F. Young}
\email[Electronic address:]{andrea@physics.ucsb.edu}
\affiliation{Department of Physics, University of California at Santa Barbara, Santa Barbara CA 93106, USA}
\date{\today}

\begin{abstract}
{\bf When bilayer graphene is rotationally faulted to an angle $\theta\approx 1.1^\circ$, theory predicts the formation of a flat electronic band and correlated insulating, superconducting, and ferromagnetic states have all been observed at partial band filling. The proximity of superconductivity to correlated insulators has suggested a close relationship between these states, reminiscent of the cuprates where superconductivity arises by doping a Mott insulator.  Here, we show that superconductivity can appear without correlated insulating states.  While both superconductivity and correlated insulating behavior are strongest near the flat band condition, superconductivity survives to larger detuning of the angle. Our observations are consistent with a ``competing phases'' picture, in which insulators and superconductivity arise from disparate mechanisms.}
 \end{abstract}
\maketitle

Strongly correlated electron systems often host dissimilar ground states vying closely for dominance.  In certain instances, doping an interaction-driven phase can give rise to further correlated phases, in a hierarchy of emergence.  Examples include the fractional quantum Hall phases that emerge doping the composite fermion sea in a partially filled Landau level\cite{jain_composite_2007} and superconducting phases that arises upon doping a Mott insulator within Hubbard models thought to apply to the cuprates\cite{lee_doping_2006}. 
However, the complexity of correlated systems allows for other possibilities; namely, different interactions can simply compete, with the winner determined by experimental details controlled by doping and material structure. 
Disentangling the origin of different phases is often hampered by the closeness of the competition, which prevents targeted control of individual parameters in the effective Hamiltonian. 

Twisted bilayer graphene (tBLG) offers an appealing platform to explore the interplay of competing states due to its high degree of tunability and panoply of apparently interaction driven ground states that break one or more system symmetries\cite{cao_correlated_2018,cao_unconventional_2018,yankowitz_tuning_2019,lu_superconductors_2019,sharpe_emergent_2019,serlin_intrinsic_2019}.  
The low energy band structure of tBLG consists of four spin- and valley-symmetry
related copies of two low-energy bands within the reduced Brillouin zone created by the moire pattern; these two bands are themselves connected by gapless Dirac points in the absence of substrate or spontaneous breaking of lattice symmetries\cite{po_origin_2018,carr_derivation_2019,koshino_maximally_2018}.  Near a twist angle of $\theta=1.1^\circ$, a resonant condition between interlayer twist angle and interlayer tunneling flattens these bands and isolates them from higher energy dispersive bands\cite{suarez_morell_flat_2010,bistritzer_moire_2011}, rendering electronic interactions relevant in determining the ground state at finite filling of these bands. A central question has been the nature of the superconductivity, which in initial reports\cite{cao_unconventional_2018,yankowitz_tuning_2019} was found to appear only in close proximity to correlated insulators near half band filling, which we denote here $\nu=\pm2$ (where $\nu=\frac{\sqrt{3}}{2}\lambda^2 n$ indicates the number of charge carriers per moire superlattice unit cell, $\lambda\approx 13$ nm the moire period and $n$ the carrier density). Initial speculation suggested that the observed superconductivity might indeed be derived from the correlated insulators, as in the cuprates. However, the observation of superconducting behavior across much wider ranges of density\cite{lu_superconductors_2019}---apparently filling the entire density range between correlated insulators at different integer $\nu$---has suggested superconductivity as a competing phase, with the correlated insulators piercing an otherwise superconducting band. A recent work\cite{stepanov_interplay_2019} has suggested that insulating states can be selectively suppressed as compared to superconductivity by enhanced screening of interactions. However, the ubiquitous observation of correlated insulating states in most superconducting devices reported in the literature continues to fuel the speculation that these two disparate phenomena may be linked, or at least share a common origin. 

\begin{figure*}[ht!]
\includegraphics[width= 7.2in]{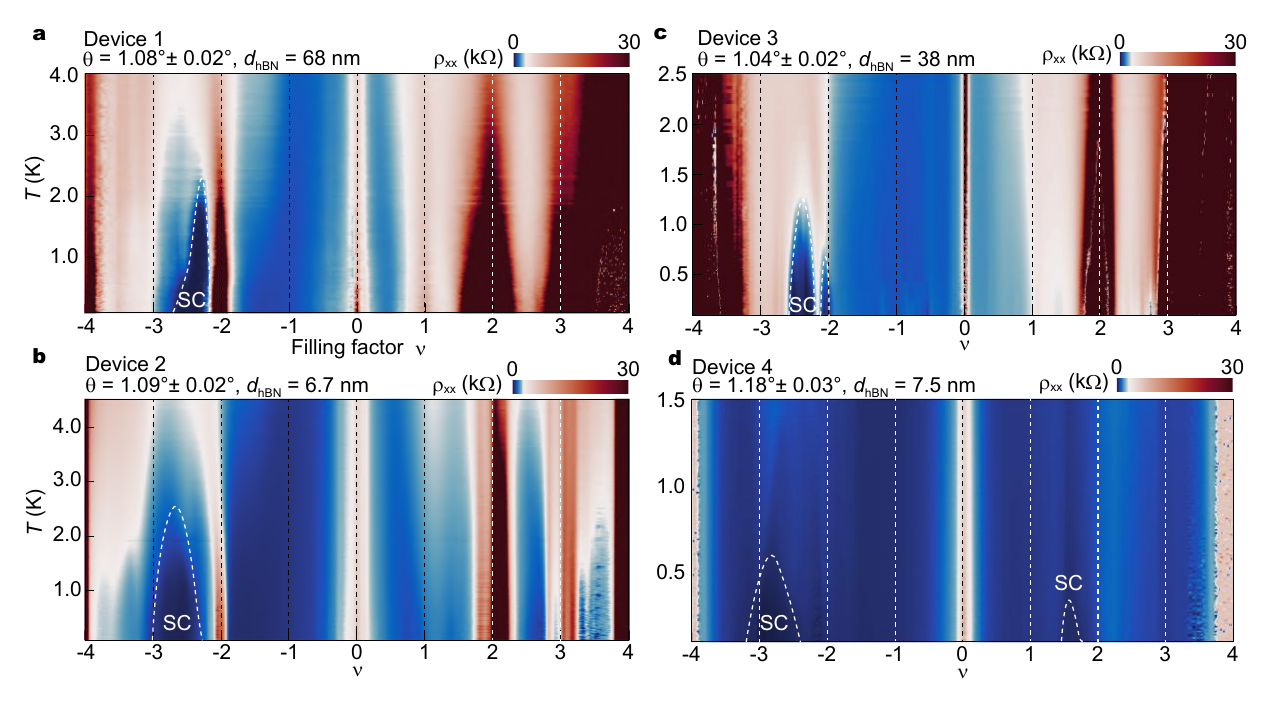}
 \caption{\textbf{Longitudinal resistance $\rho_\mathrm{xx}$ as a function of temperature $T$ and filling factor $\nu$ in four devices.} \textbf{a}, \textbf{b}, \textbf{c} and \textbf{d} show data for Devices 1, 2, 3 and 4, respectively. $d_\mathrm{hBN}$ is the thickness of the hBN gate dielectric separatig the tBLG from a graphite gate. The dashed lines mark integer $\nu$. 
 The dashed curves around the superconducting dome shows rough boundary between superconducting and normal state, as determined by a resistance drop of 50\% relative to the normal state state.}
\label{fig:1}
\end{figure*}

Here, we report superconducting behavior in tBLG devices decoupled from the appearance of correlated insulators, strongly suggesting disparate origins for the superconductivity and correlated insulating behavior.
Figure \ref{fig:1} shows resistivity maps for four devices near the flat band condition as a function of temperature and carrier density with the later tuned across the entire range of the low energy bands,  $-4<\nu<4$. All devices show robust signatures of superconductivity near $\nu \approx -2$ (see also Figs. \ref{fig:linecutRvsN}, \ref{fig:SCdomes} and \ref{fig:linecutRT}), manifesting as a transition to a low-resistivity state at low temperatures, large non-linearity at low applied currents (a critical current), and phase coherent effects reminiscent of Fraunhoffer patterns in Josephson junctions. All tBLG devices used in this study are fabricated by a ``cut-and-stack'' technique (see Methods and Fig. \ref{fig:stackprocess} for details).  
In contrast to ``tear-and-stack''\cite{kim_van_2016, cao_superlattice-induced_2016}, in which a monolayer is torn by an encapsulating hBN flake, we first cut the monolayer using an atomic force microscope (AFM)\cite{chen_evidence_2019} in order to prevent unintentional strain developing during the stacking process. This fabrication technique reproducibly leads to devices with twist angle inhomogeneity of $\lesssim 0.03^\circ$ on $\sim$ 10 $\mu$m length scale, as quantified by two terminal conductance measurements across different contacts (Fig. \ref{fig:contacts}).
The devices further employ a bottom graphite or graphene gate to reduce charge inhomogeneity\cite{yankowitz_tuning_2019,lu_superconductors_2019}.
Devices 1, 2 and 5 (shown in Figs. \ref{fig:1}a, b and \ref{fig:SCdomes}e, respectively) are closest to the flat band condition with $\theta = 1.08^\circ$, 1.09$^\circ$, and 1.12$^\circ$, respectively. All show superconductivity, correlated insulating states at $\nu = -2$, as well as additional insulating states or resistivity peaks at $\nu= +2$ and $+3$  consistent with previous observations in homogeneous tBLG under high\cite{yankowitz_tuning_2019} and ambient pressure\cite{lu_superconductors_2019}. In contrast, Devices 3 and 4, which are both farther from the flat band condition in angle at $\theta$=1.04$^\circ$ and 1.18$^\circ$, respectively, show superconductivity despite the absence of an insulating state at $\nu=-2$ (for Device 3) and total absence of any correlated insulating states at all (for Device 4). 

\begin{figure*}[ht!]
\includegraphics[width= 7.2in]{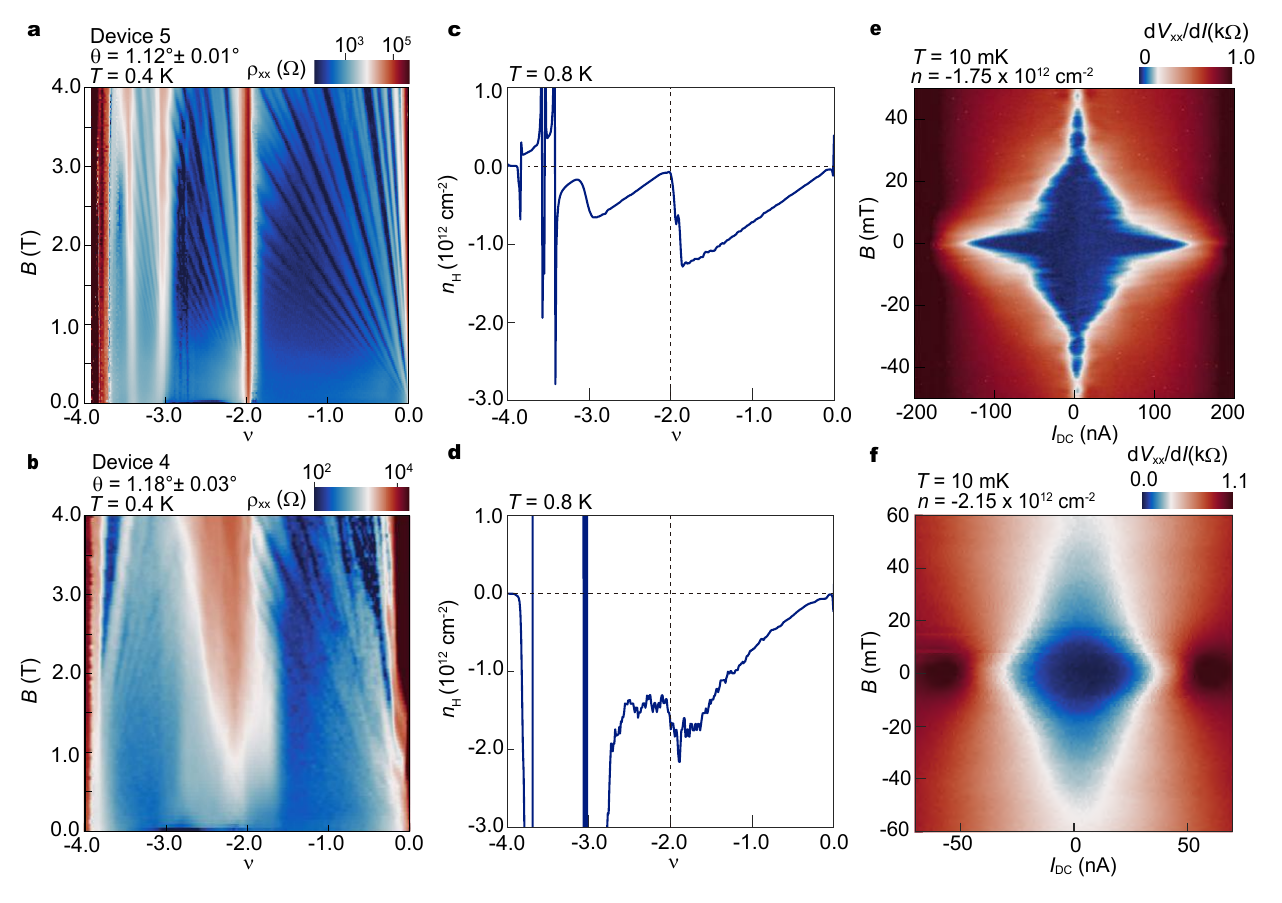}
 \caption{\textbf{Distinct transport behavior between Devices 4 and 5.}
 \textbf{a}, \textbf{b},  Landau-fan diagram near $\nu= -2$ at 0.4 K in Devices 5 (a) and 4 (b). Device 4 shows separate sequences of quantum oscillations originating from $\nu=-2$ as well as $\nu=-3$, while in Device 4 the only visible quantum oscillations originat at charge neutrality or full band filling.
\textbf{c}, \textbf{d},  Hall density ($n_\mathrm{H}$) as a function of $\nu$ Devices 5 (c) and 4 (d) at 0.8 K. The Hall density in Device 5 resets to zero at $\nu = -2$, indicating the formation of a new, small Fermi surface. No such effect is observed in Device 4. 
 \textbf{e}, \textbf{f}, Critical current as a function of magnetic field in Devices 5 (e) and 4 (f), measured at electron densities of  $-1.75\times10^{12}$ cm$^{-2}$ and $-2.15\times10^{12}$ cm$^{-2}$, respectively.  Measurements were performed at a nominal temperature of 10 mK.
}
\label{fig:2}
\end{figure*}

The contrasting behavior observed between Devices 1-2 and 3-4 raises the question  of the role of disorder, particularly inhomogeneities in $\theta$ which may vary from sample to sample and obscure the intrinsic phenomenology of tBLG.  For example, previous studies of twisted bilayer graphene have found that in the low temperature limit, superconductivity may obtain even at exactly $\nu=-2$\cite{cao_unconventional_2018,yankowitz_tuning_2019}, a finding attributed to percolation of superconducting domains due to variations in the local $\nu$ across the sample. The absence of an insulating state at $\nu=-2$ in the devices presented here appears instead to be intrinsic, based on several observations. 
First, devices with and without a correlated insulator at $\nu=-2$ show distinct behavior in finite magnetic field (Figs. \ref{fig:2}a and b). 
Fig. \ref{fig:2}a shows magnetoresistance of Device 5. A sequence of quantum oscillations, with an apparent degeneracy of 2, is clearly visible originating at $\nu=-2$, indicating the formation of a new Fermi surface associated with this insulating state. In contrast, for Devices 3 and 4 (Figs. \ref{fig:2}b and  \ref{fig:landaufan_wg26}), we observe no strong quantum oscillations originating from $\nu = -2$. In addition, the apparent resetting of the Hall density typically concomitant with the appearance of a correlated insulator at $\nu = -2$ (Fig. \ref{fig:2}c) as is observed in Device 5  (see also Figs. \ref{fig:Hall}a and b), is suppressed in Device 3 (Fig. \ref{fig:Hall}c) and nearly completely absence in Device 4 (Fig. \ref{fig:2}d). In Device 4, shown in Fig. \ref{fig:2}b, the neighborhood of $\nu=-2$ shows only a strong magnetoresistance over a broad range of filling factor, devoid of oscillations but bracketed on either side by quantum oscillations originating from $\nu=0$ and $\nu=4$. The distinct behavior of quantum oscillations and Hall effect in Devices 3 and 4 support the absence of significant interaction-induced Fermi surface reconstruction near $\nu = -2$ in detuned devices as an intrinsic property. 

While the quantum oscillations show highly contrasting behavior between flat band and detuned devices, superconducting states appear to be phenomenologically similar to each other. Figures \ref{fig:2}e and f show d$V$/d$I$ versus magnetic field ($B$) and DC current ($I_\mathrm{DC}$) at 10 mK in Devices 5 and 4 at density of $-1.75\times10^{12}$ and $-2.15\times10^{12}$ cm$^{-2}$, respectively; similar data for Devices 1, 2 and 3 is shown in Fig. \ref{fig:otherfraunhofer}. 
All devices show an apparent critical current, weakly modulated by the applied magnetic field. The period of the oscillations $\Delta B$ varies from around 3$-$8 mT (see Figs. \ref{fig:2}e and \ref{fig:fraunlinecut}), indicating an effective junction area of S $\sim$ 0.25$-$0.67 $\mu$m$^2$ , using S = $\Phi\Delta B$, where $\Phi$ = $h/2e$ is the superconducting flux quantum, $h$ is Planck’s constant, and $e$ is the charge of the electron. In prior work, these Fraunhofer-like oscillations have been attributed to the presence of Josephson junctions consisting of small metallic or insulating domains within the device\cite{cao_unconventional_2018,yankowitz_tuning_2019,lu_superconductors_2019}. 
Crucially, the Fraunhofer-like quantum interference pattern provides independent evidence of  superconductivity beyond the vanishing low temperature resistance. This is critical for 2D superconductors, where the absence of detectable Meissner effect deprives experimentalists of the primary indicator of superconductivity in 3D materials. This is particularly important in high quality electronic systems such as graphene where ballistic transport can easily cause measured resistance to drop to zero\cite{mayorov_direct_2011}.

\begin{figure*}[ht!]
\includegraphics[width= 7.2in]{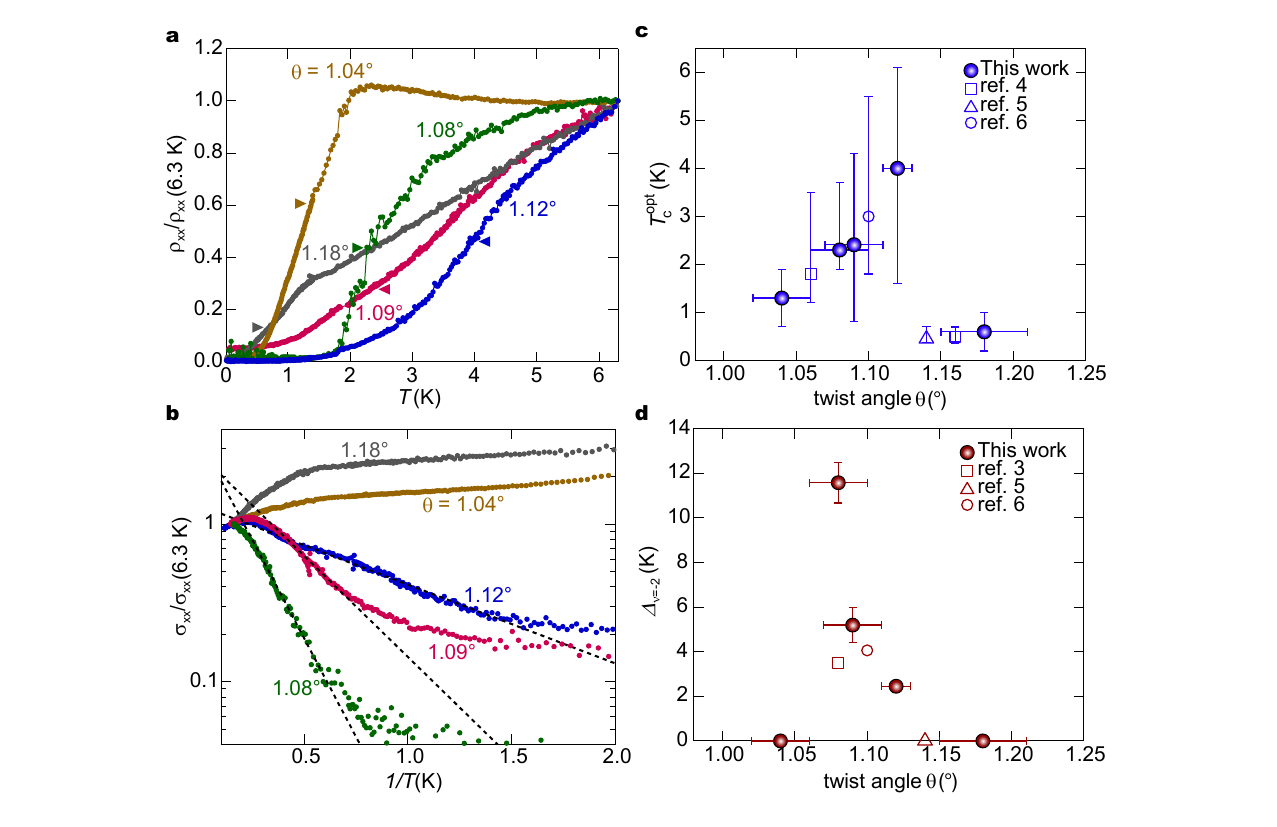}
 \caption{\textbf{Superconducting and correlated insulating states near $\nu=-2$ filling.}
\textbf{a}, $\rho_\mathrm{xx}(T)$ normalized by its value at 6.3 K measured at the optimal doping for superconductivity for Devices 1-5. Arrows indicate 50\% of the normal state resistance, which we take to define $T_\mathrm{c}^\mathrm{opt}$.
 \textbf{b}, Four terminal conductivity $\sigma_\mathrm{xx}(T)$ normalized by $\sigma_\mathrm{xx}$(6.3 K) at $\nu = -2$. The black dashed lines show fits to an Arrhenius law, $\sigma_\mathrm{xx}\sim \exp(-2\Delta /T)$.
 \textbf{c}, $T_\mathrm{c}^\mathrm{opt}$ as a function of twist angles. $T_\mathrm{c}^\mathrm{opt}$ is defined by the midpoint of the resistive transition, as shown in \textbf{a}. The upper and lower limit of error bars are 90\% and 10\% of the resistive transition, respectively. Open squares, circle and triangles are taken from published data in References \cite{cao_unconventional_2018,yankowitz_tuning_2019,lu_superconductors_2019}.
 \textbf{d}, The activation gap at $\nu = -2$ as a function of twist angle.  Error bars in the gaps represent the uncertainty arising from determining the linear (thermally activated) regime for the fit. Open square, circle and triangles are taken from published data in References \cite{cao_correlated_2018,yankowitz_tuning_2019,lu_superconductors_2019}}
\label{fig:3}
\end{figure*}

To quantitatively compare transport data across devices, Fig. \ref{fig:3}a shows the temperature dependent resistance at the optimal doping for superconductivity normalized to the its value at 6.3 K, while Fig. \ref{fig:3}b shows Arrhenius plots of the four terminal conductivity $\sigma_\mathrm{xx}$ at $\nu=-2$. Most of the superconducting transitions show a broad characteristic as compared to other 2D superconductors \cite{yazdani_superconducting-insulating_1995, reyren_superconducting_2007}, complicating assignment of a critical temperature; we 
define $T_\mathrm{c}^\mathrm{opt}$ as the temperature where $\rho_\mathrm{xx}$ is 50\% of the normal state resistance, indicated by triangles in Fig. \ref{fig:3}a.
Devices 1, 2 and 5 show relatively high $T_\mathrm{c}$, comparable to the value reported in previous high-homogeneity 1.10$^\circ$ tBLG\cite{lu_superconductors_2019}. They also show activated behavior of the conductivity at $\nu=-2$, $\sigma_{xx}\sim \exp({-\Delta/T})$, allowing the gap $\Delta_\mathrm{\nu = -2}$ to be determined from the experimental data.  
In Devices 3 and 4 ($\theta = 1.04^\circ$ and $1.18^\circ$), in contrast, there is no well-developed thermal activation gap at $\nu = -2$.  Nevertheless, the devices show a clear superconducting transition, with a $T_\mathrm{c}$ slightly lower than in the flat band devices. We summarize these observations in Figs. \ref{fig:3}c and d. Both $T_\mathrm{c}^\mathrm{opt}$ and  $\Delta_\mathrm{\nu = -2}$ trace out a dome-shaped behavior as a function of twist angle, reaching a  maximum near 1.1$^\circ$.  However, finite $\Delta_\mathrm{\nu = -2}$ appears over a measurably narrower domain of $\theta$ as compared to the domain of finite $T_\mathrm{c}^\mathrm{opt}$.

Twisted bilayer graphene is a complex system in which the low temperature phase diagram is highly sensitive to a number of independent experimental parameters. 
As is evident in our data and consistent with theoretical modeling\cite{bistritzer_moire_2011}, the strength of interactions relative to bandwidth is highly dependent on twist angle, with detuning from 1.1$^\circ$ significantly reducing the role of interactions.  
Screening of electronic interactions may also play a role\cite{goodwin_critical_2019}. In particular, Devices 2 and 4 both feature graphite gates that are closer to the tBLG than the moire wavelength, which is expected to significantly screen Coulomb interactions\cite{goodwin_critical_2019,pizarro_internal_2019} as argued in a recent experimental study that also finds superconductivity without insulating states in tBLG\cite{stepanov_interplay_2019}. 
Disambiguating these effects precisely is complicated by the effect of twist angles between the tBLG layers and hBN encapsulants, as well as strain.  Both of these effects (and presumably other, even more subtle sample details) are not controlled experimentally, and likely vary between devices. Unlike the twist angle or gate screening they are difficult quantify, and thus prevent carefully controlled experiments in which a single parameter is varied.  The phase diagram of Figs. \ref{fig:3}c-d thus represents only a coarse cut through the multidimensional parameter space governing tBLG physics. 

Nevertheless, the observation of superconductivity in the absence of correlated insulating states constitutes powerful evidence that superconductivity arises independently.  While more exotic scenarios cannot be ruled out definitively, a simple picture is that the superconductivity arises through the usual electron-phonon coupling\cite{wu_theory_2018,lian_twisted_2019,peltonen_mean-field_2018, angeli_valley_2019}. As a Fermi surface instability, superconductivity is sensitive to the density of states at the Fermi level rather than the total bandwidth. In contrast, the correlated insulating states are likely to break spin or valley symmetry, requiring a complete polarization of the band at an energy cost equivalent to the total bandwidth, as follows from a Stoner criterion.  Correlated insulators thus become favored only when the bandwidth becomes exceptionally narrow, while superconductivity may be favored even far from the flat band condition, as the peak density of states remains high.  Near 1.1$^\circ$, correlated insulators win the energetic competition at commensurate fillings, rearranging the electronic bands in the process and leading to the observation of superconductivity at diverse filling factors that vary from device to device\cite{cao_unconventional_2018,yankowitz_tuning_2019,lu_superconductors_2019}.  In this picture, tBLG resembles alkali-doped $C_{60}$\cite{ramirez_superconductivity_2015}, in which BCS superconductivity or correlated insulator physics obtain at different dopings.   


\let\oldaddcontentsline\addcontentsline
\renewcommand{\addcontentsline}[3]{}
\section*{Methods}
\vspace{-12pt}

\noindent\textbf{Device fabrication}\\ The tBLG used in this study were fabricated using a dry-transfer and ``cut-and-stack'' technique (Fig. \ref{fig:stackprocess}) instead of the ``tear-and-stack'' technique\cite{kim_van_2016, cao_superlattice-induced_2016}. Prior to stacking, we first cut graphene into two pieces using AFM\cite{chen_evidence_2019} (Fig. \ref{fig:stackprocess}a) to prevent the unintentional strain in tearing graphene. We used a poly(bisphenol A carbonate) (PC)/polydimethylsiloxane (PDMS) stamp mounted on a glass slide to pick up a hBN flake (typically 30$-$50 nm) at 90$-$110$^\circ$C, and carefully pick up the 1st half of pre-cut graphene piece, rotate and pick up again the 2nd half of graphene piece in series at 25 $^\circ$C using this hBN flake (Fig. \ref{fig:stackprocess}b for details). Here we rotated graphene pieces manually by a twist angle of about 1.2$^\circ-$1.3$^\circ$. Finally, the 3-layer stack (hBN-tBLG) is transferred onto another stack for the bottom gate part (hBN-graphite or graphene gate), which is prepared in advance by the same dry transfer process and cleaned by the typical solvent wash using chloroform, acetone, methanol and IPA followed by vacuum annealing (400$^\circ$C for 8 hours) to remove the residue of PC film on the hBN surface. We did neither squeeze the bubbles nor perform any heat annealing after the stack is completed to prevent the relaxation of tBLG. We selected a bubble free region as a channel area to prevent inhomogeneity. Electrical connections to the tBLG were made by CHF$_3$/O$_3$ etching and deposition of the Cr/Pd/Au (2/15/180 nm) metal edge-contacts\cite{wang_one-dimensional_2013}. Following this process, we got 5 superconducting tBLG devices (presented in this paper) of about 20$-$25 stacks.

\noindent\textbf{Transport measurements}\\All transport measurements in this study were carried out in a dilution refrigerator (Bluefors LD400) with a base temperature of 10 mK, which is equipped with a 14 T superconducting magnet and heavy RF and audio frequency filtering with a cutoff frequency of $\sim$ 10 kHz. The temperature dependent measurements were done by controlling the temperature using a heater mounted on a mixing chamber plate. Standard low frequency lock-in techniques with Stanford Research SR860 amplifiers were used to measure the resistance $\rho_\mathrm{xx}$ and $\rho_\mathrm{xy}$ with an excitation current of 1$-$2 nA at a frequency of 17.777 Hz.  

\noindent\textbf{Twist angle determination}\\
The twist angle $\theta$ is determined from the values of charge carrier density at which the insulating states at $\pm n_s$ are observed, following $n_s = 8 \theta^2/\sqrt{3}a^2$ , where $a$= 0.246 nm is the lattice constant of graphene. The values of $\pm n_s$ are determined from the sequence of quantum oscillations in a magnetic field that project to $\pm n_s$ or $\pm n_s/2$ for devices.
\vspace{5pt}
\let\addcontentsline\oldaddcontentsline


\let\oldaddcontentsline\addcontentsline
\renewcommand{\addcontentsline}[3]{}

\section*{Acknowledgments}
\vspace{-12pt}
\noindent
We thanks M. Yankowitz and D.K. Efetov for discussions and D.K. Efetov for sharing unpublished experimental data. This work was primarily supported by the ARO under W911NF-17-1-0323.
Y.S. acknowledges the support of the Elings Prize Fellowship.
K.W. and T.T. acknowledge support from the Elemental Strategy Initiative conducted by the MEXT, Japan and the CREST(JPMJCR15F3), JST.
A.F.Y. acknowledges the support of the David and Lucille Packard Foundation and the Alfred P. Sloan foundation.
\let\addcontentsline\oldaddcontentsline

\let\oldaddcontentsline\addcontentsline
\renewcommand{\addcontentsline}[3]{}
\section*{Author Contributions}
\vspace{-12pt}
\noindent
Y.S. and J.G. fabricated tBLG devices. Y.S. performed the measurements and analyzed the data. Y.S. and A.F.Y. wrote the manuscript. T.T. and K.W. grew the hBN crystals.
\let\addcontentsline\oldaddcontentsline

\let\oldaddcontentsline\addcontentsline
\renewcommand{\addcontentsline}[3]{}
\section*{Competing interests}
\vspace{-12pt}
\noindent
The authors declare no competing financial interests.
\let\addcontentsline\oldaddcontentsline

\let\oldaddcontentsline\addcontentsline
\renewcommand{\addcontentsline}[3]{}
\bibliographystyle{unsrt}
\bibliography{references}
\let\addcontentsline\oldaddcontentsline

\clearpage

\pagebreak
\widetext
\begin{center}
\textbf{\large Supplementary Materials for \\Decoupling superconductivity and correlated insulators in twisted bilayer graphene}\\
\vspace{10pt}
Yu Saito, Jingyuan Ge, Kenji Watanabe, Takashi Taniguchi and Andrea F. Young\\
\vspace{10pt}
Correspondence to: andrea@physics.ucsb.edu\\
\vspace{10pt}
\end{center}
\renewcommand{\thefigure}{S\arabic{figure}}
\renewcommand{\thesubsection}{S\arabic{subsection}}
\setcounter{secnumdepth}{2}
\renewcommand{\theequation}{S\arabic{equation}}
\renewcommand{\thetable}{S\arabic{table}}
\setcounter{figure}{0}
\setcounter{equation}{0}
\onecolumngrid

\begin{figure*}[ht!]
\includegraphics[width= 7.2in]{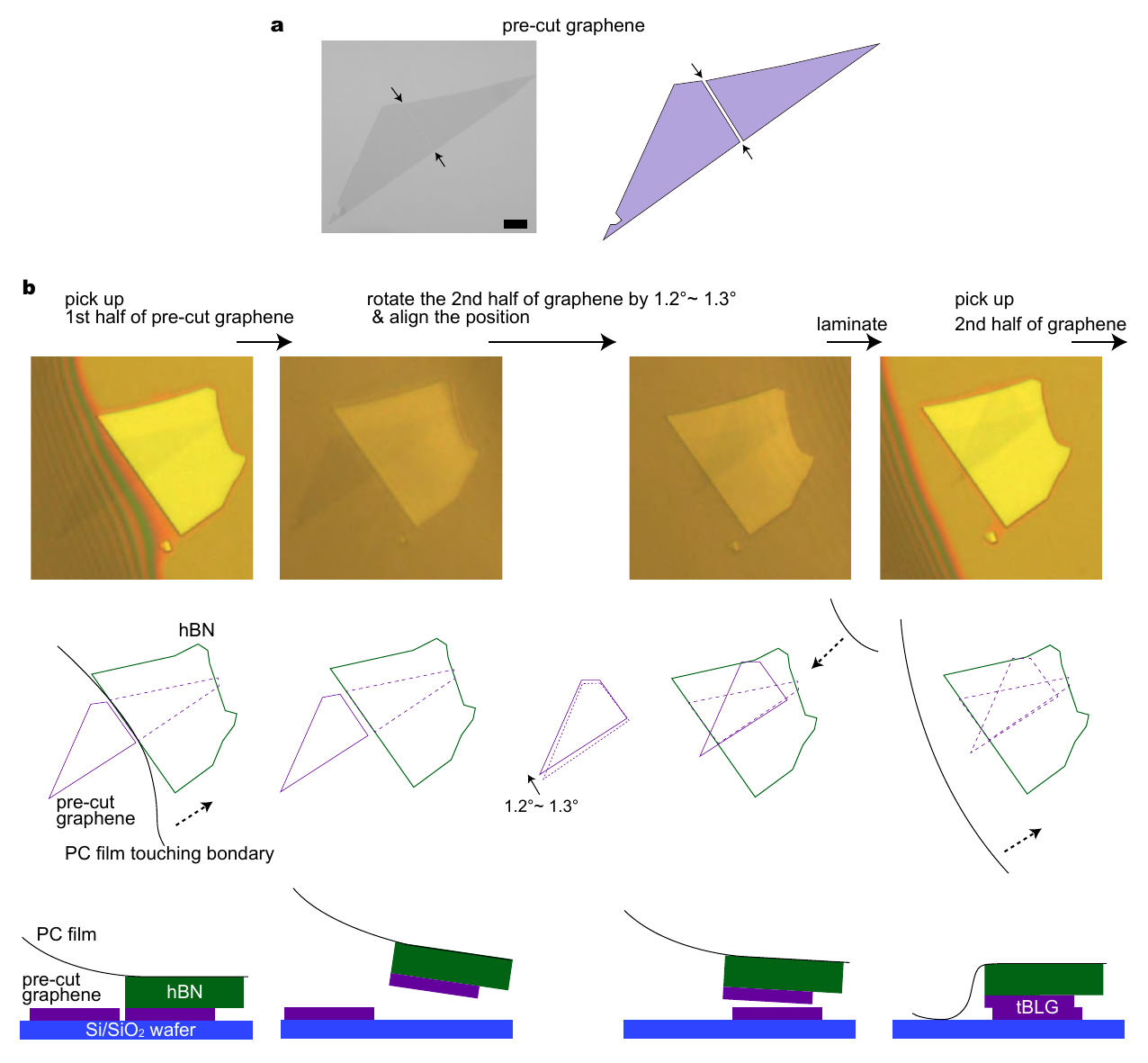}
\caption{\textbf{``Cut-and-stack'' fabrication technique for tBLG.} \textbf{a}, Optical microscope and schematic images of graphene pre-cutting.  A graphene flake is cut into two or more pieces by AFM\cite{chen_evidence_2019}. Arrows show pre-cutting line. Scale bar in left images is 10 $\mu$m. 
\textbf{b}, Process flow for tBLG fabrication. All stacking is done at 25$^\circ$C. First, we laminate the 1st half piece of pre-cut graphene by carefully aligning the edge of hBN with the cutting line and picking it up with the hBN flake. After that, we rotate the 2nd half-piece of graphene by about 1.2$^\circ-$1.3$^\circ$ and laminate/pick it up. Dashed arrows show the direction of motion of boundary between adhered and non-adhered PC film.}
\label{fig:stackprocess}
\end{figure*}

\begin{figure*}[ht!]
\includegraphics[width= 7.2in]{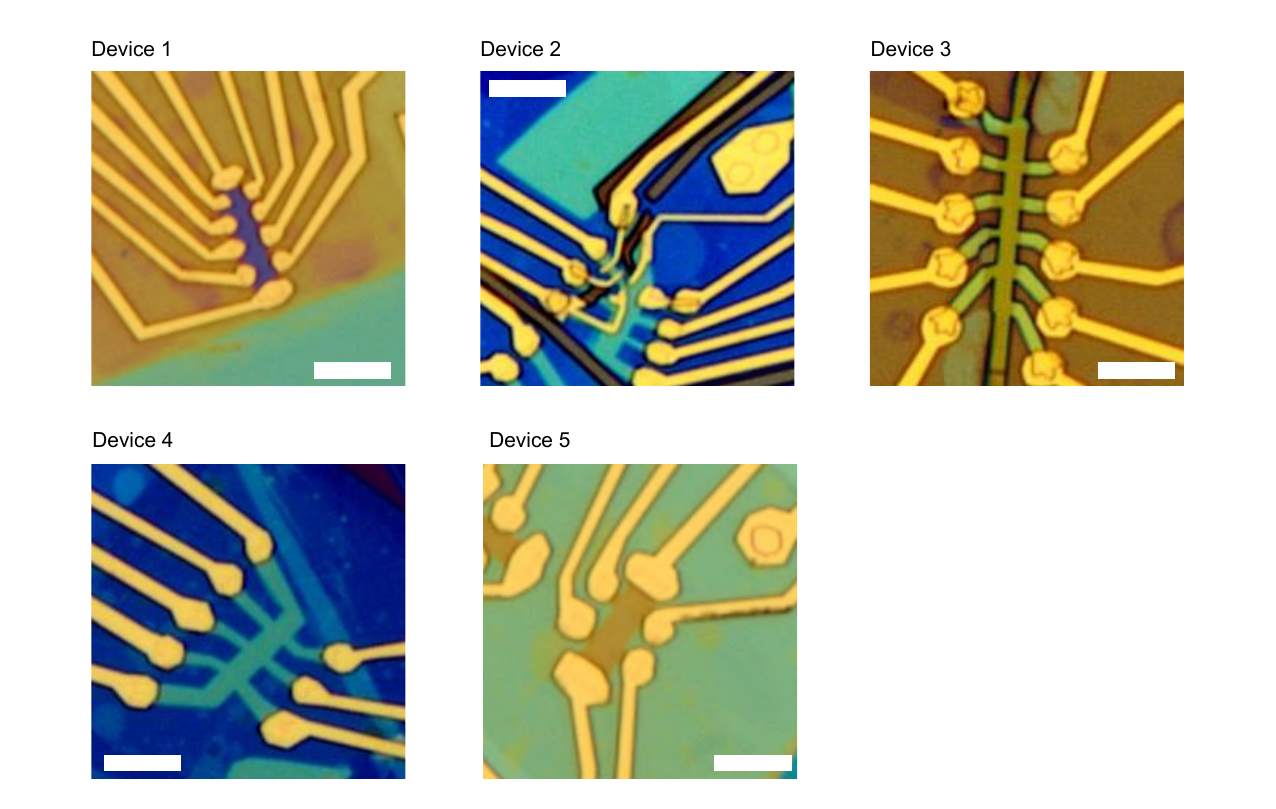}
\caption{\textbf{Optical microscope images of devices used in this study.} Scale bars equal 5 $\mu$m in all images.}
\label{fig:devices}
\end{figure*}

\begin{figure*}[ht!]
\includegraphics[width= 7.2in]{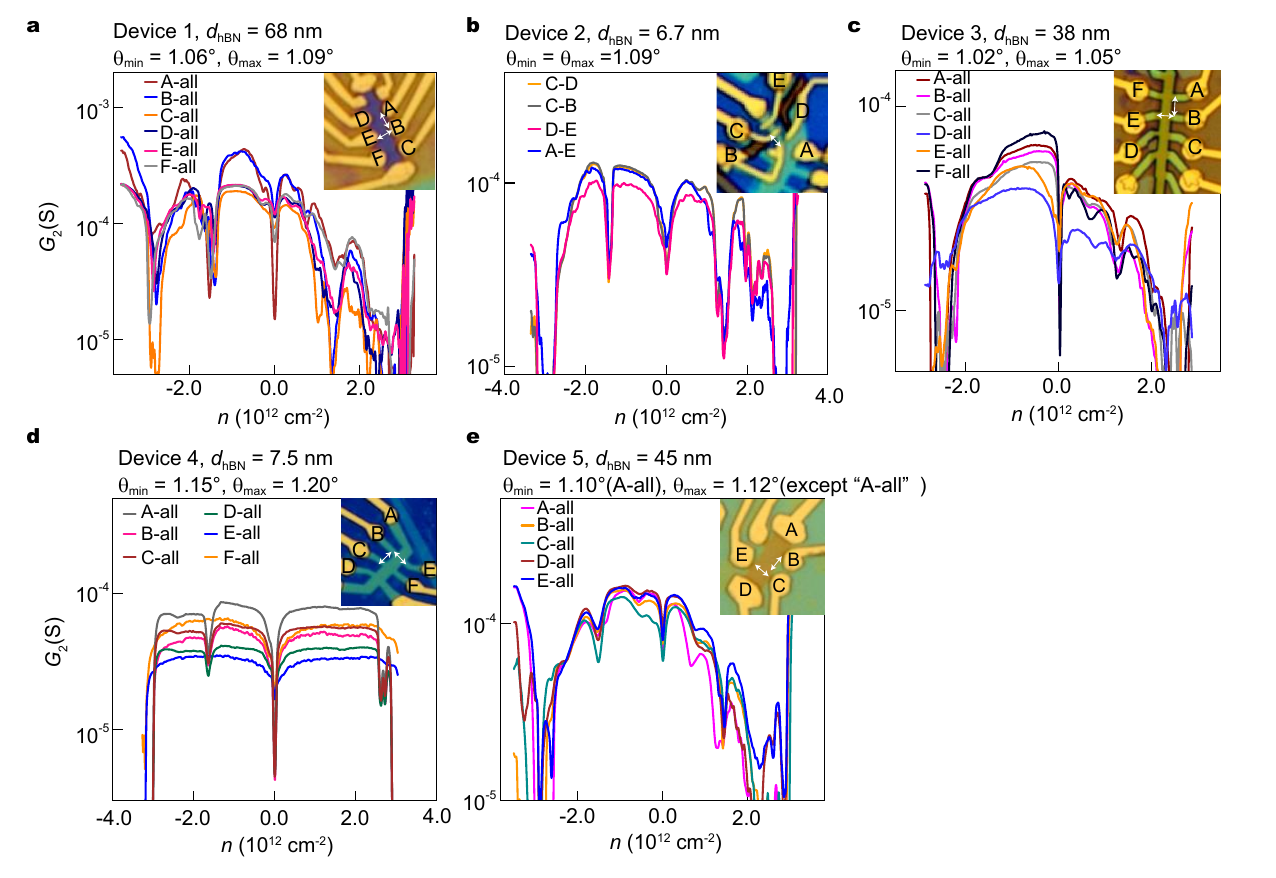}
\caption{\textbf{Two terminal conductance across multiple contacts in all five measured tBLG devices.}
Measurements were performed at 0.8 K for Devices 1-4 and 4 K for Device 5. White arrows show four terminal contacts used for $\rho_\mathrm{xx}$ and $\rho_\mathrm{xy}$. $\theta_\mathrm{max}$ and $\theta_\mathrm{min}$ are the largest and smallest twist angles calculated from superlattice peaks and $\nu = \pm 2$ peaks. The left side contacts (A, B, C and D) of Device 4 shows 1.15$^\circ$ while the right side (E and F) is 1.20$^\circ$. }
\label{fig:contacts}
\end{figure*}

\begin{figure*}[ht!]
\includegraphics[width= 7.2in]{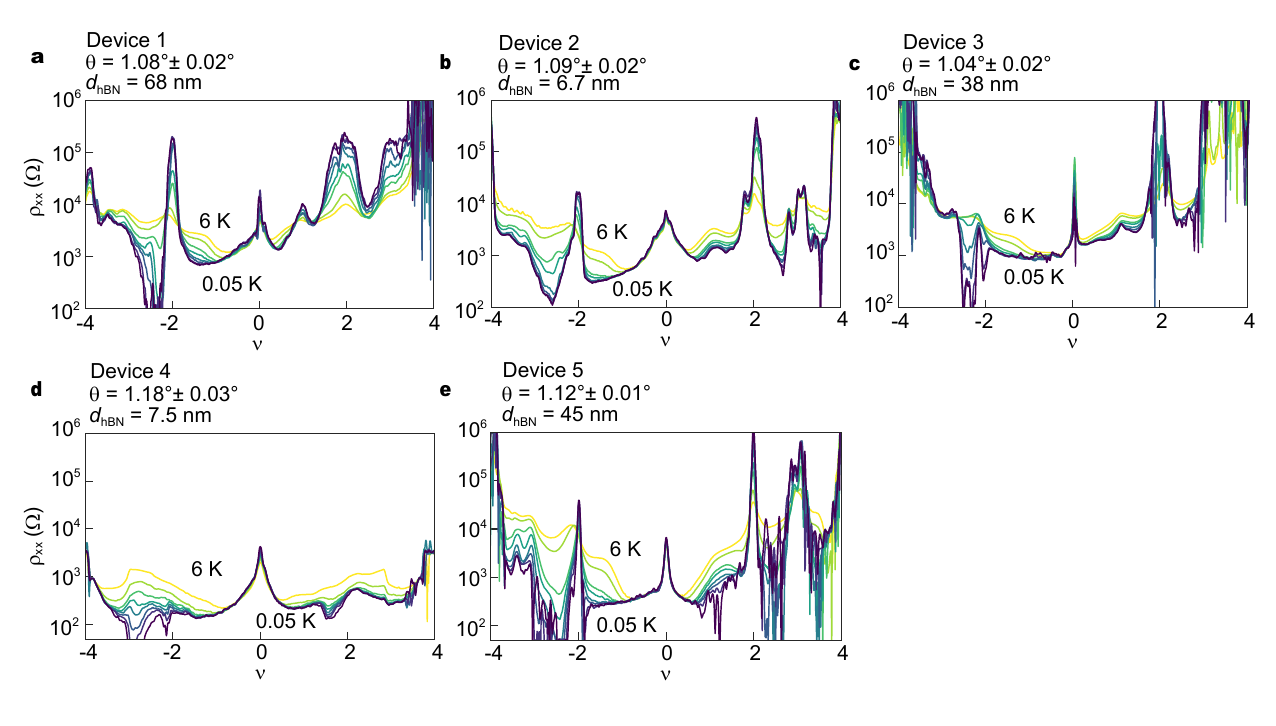}
\caption{\textbf{Line cuts of $\rho_\mathrm{xx}$ versus filling factor $\nu$ between 6 K and 50 mK.} 
The curves are at 6, 4, 2, 1.5, 0.9, 0.6, 0.3 and 0.05 K for Devices 1, 2, 3 and 5, and 6, 3, 1.5, 0.6, 0.4, 0.3 and 0.05 K for Device 4.}
\label{fig:linecutRvsN}
\end{figure*}

\begin{figure*}[ht!]
\includegraphics[width= 7.2in]{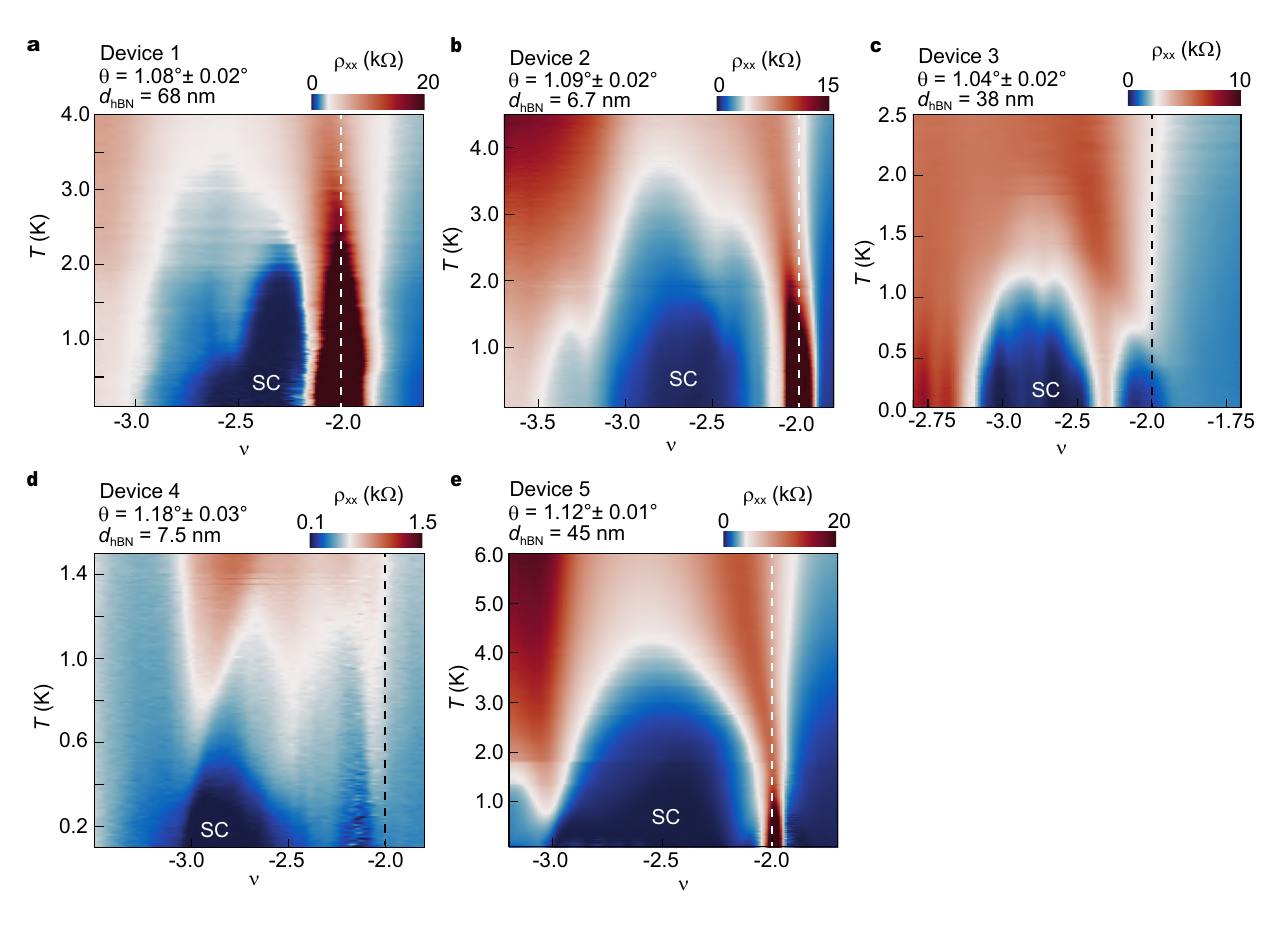}
\caption{\textbf{Detail of 2D map around a superconducting dome in each device.} 
Dashed lines show $\nu=-2$ filling. The superconducting phase in Device 3 is divided by a weak resistive state around $\nu = -2-\delta$, which does not match the density of the $\nu = -$2 filling, estimated from the strong resistive states at $\nu = -4, 0,  2$ and $4$.}
\label{fig:SCdomes}
\end{figure*}

\begin{figure*}[ht!]
\includegraphics[width= 7.2in]{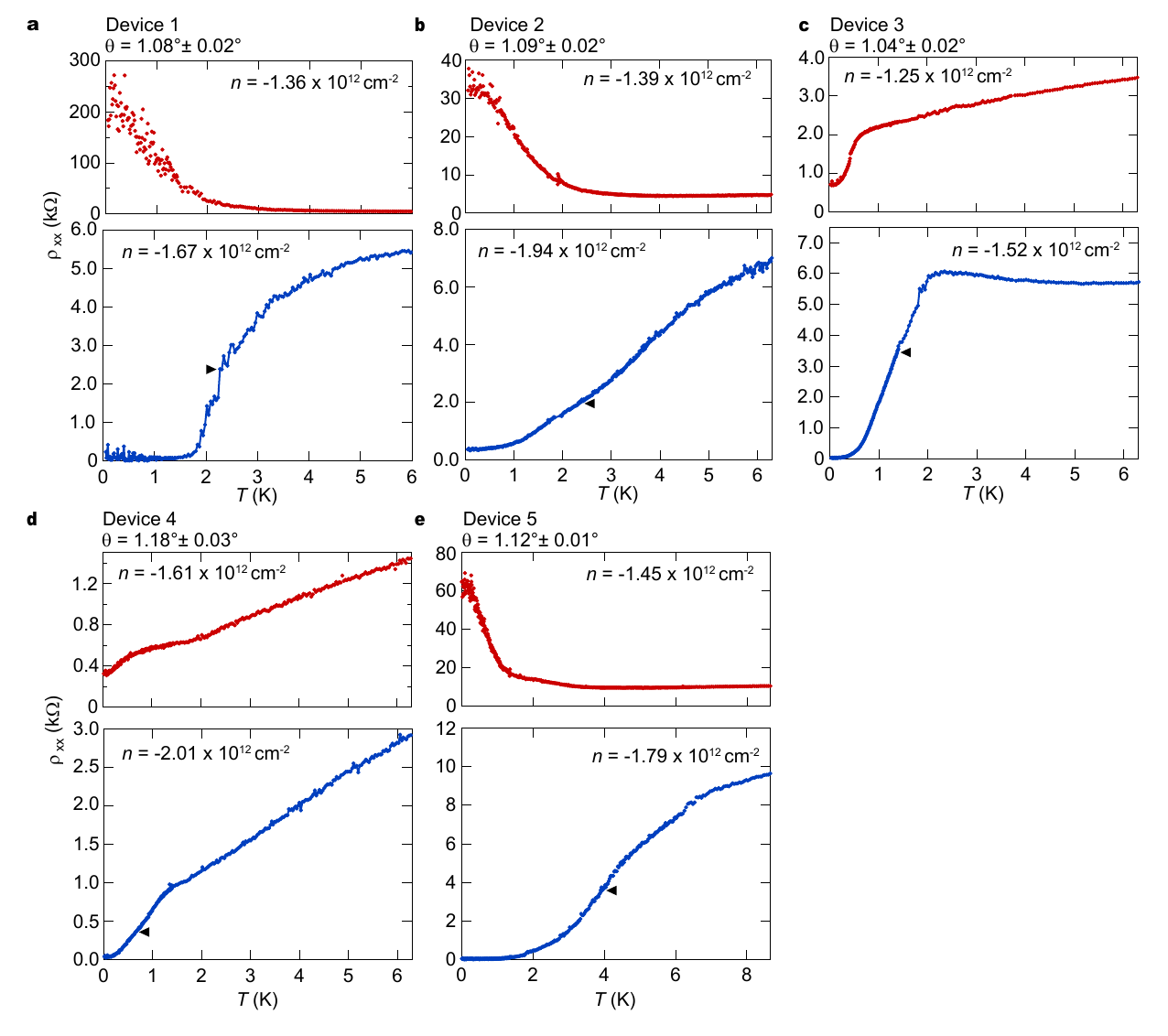}
\caption{\textbf{$\rho_\mathrm{xx}(T)$at optimal doping for superconductivity (blue curves) and $\nu=-2$ filling (red curves).} The black triangles show indicate 50\% deviation from normal state resistance.}
\label{fig:linecutRT}
\end{figure*}

\begin{figure*}[ht!]
\includegraphics[width= 7.2in]{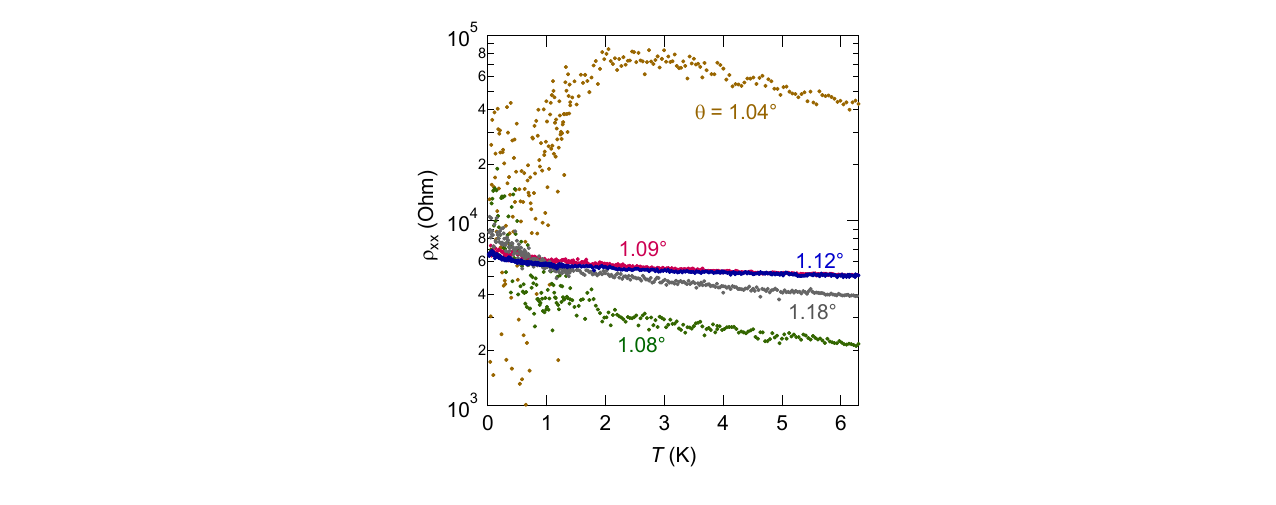}
\caption{\textbf{$\rho_\mathrm{xx}(T)$ at charge neutrality point in the five devices measured.}}
\label{fig:CNPRT}
\end{figure*}

\begin{figure*}[ht!]
\includegraphics[width= 7.2in]{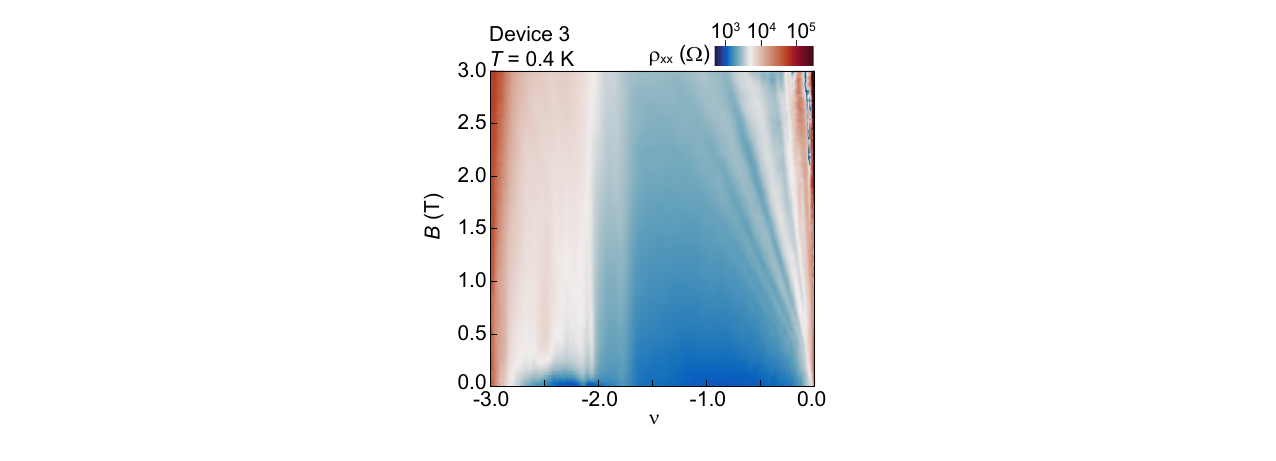}
\caption{\textbf{Magnetoresistance at negative densities in Device 3 at 0.4 K}.} 
\label{fig:landaufan_wg26}
\end{figure*}

\begin{figure*}[ht!]
\includegraphics[width= 7.2in]{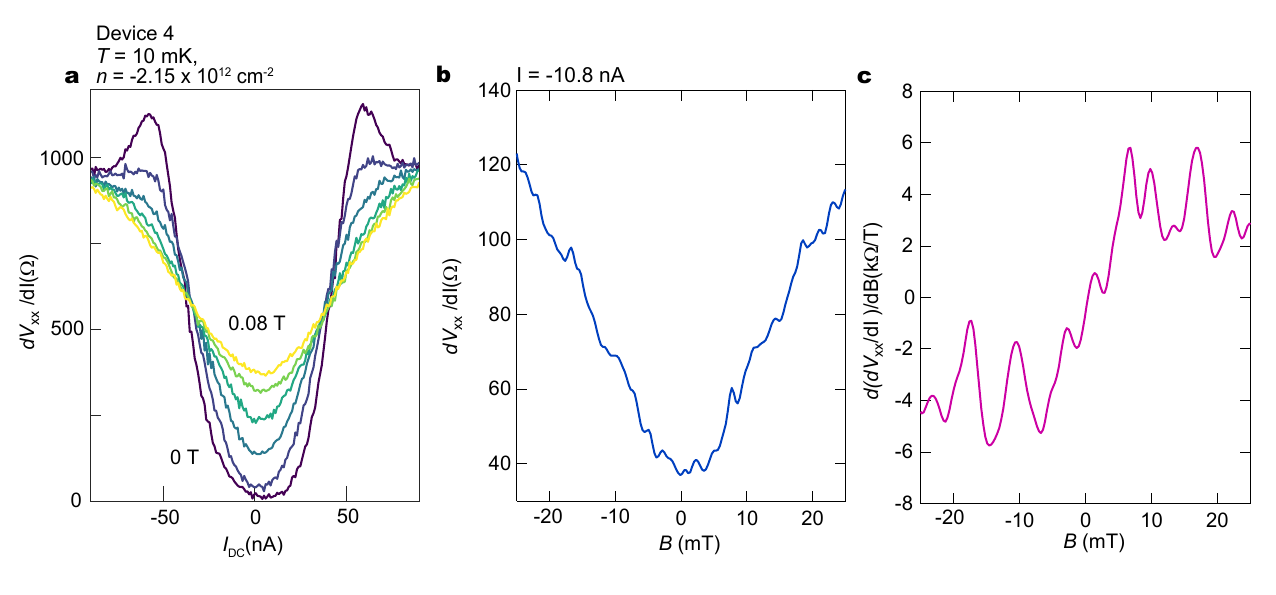}
\caption{\textbf{Fraunhofer-like interferences in Device 4}
Differential resistance $dV_\mathrm{xx}/dI$ as a function of DC current ({\bf a}) and magetic field ({\bf b}). The latter is measured for $-10.8$ nA DC current.  Field derivative of differential resistance $d^2V_\mathrm{xx}/dIdB$ is plotted as a function of $B$ in {\bf c}, and shows periodic oscillations reminiscent of a Fraunhofer patterns. } 
\label{fig:fraunlinecut}
\end{figure*}

\begin{figure*}[ht!]
\includegraphics[width= 7.2in]{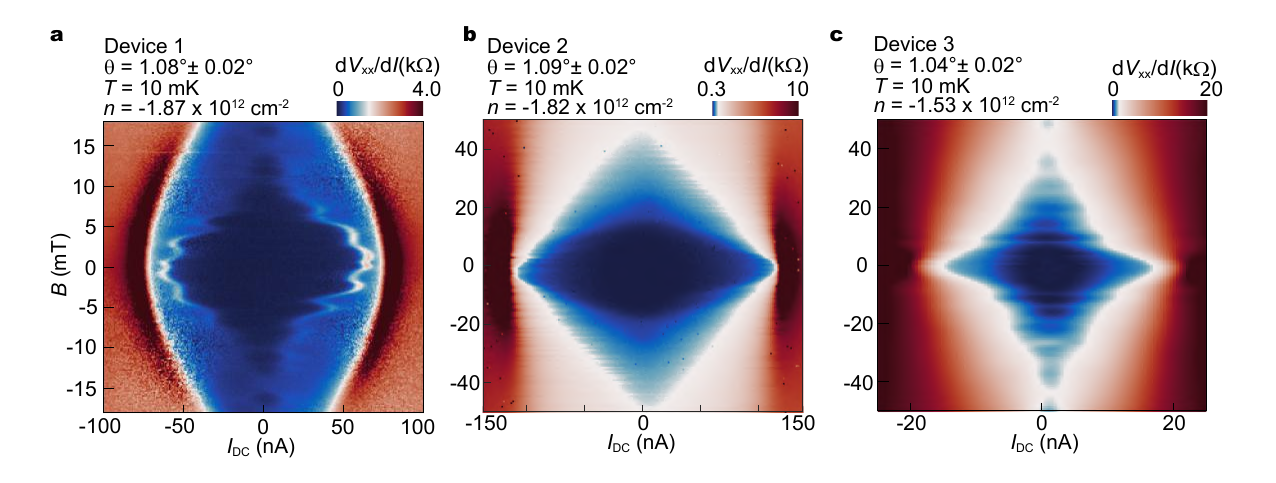}
\caption{\textbf{Differential resistance as a function of DC current and magnetic field in Devices 1, 2 and 3 at 10 mK.} } 
\label{fig:otherfraunhofer}
\end{figure*}

\begin{figure*}[ht!]
\includegraphics[width= 7.2in]{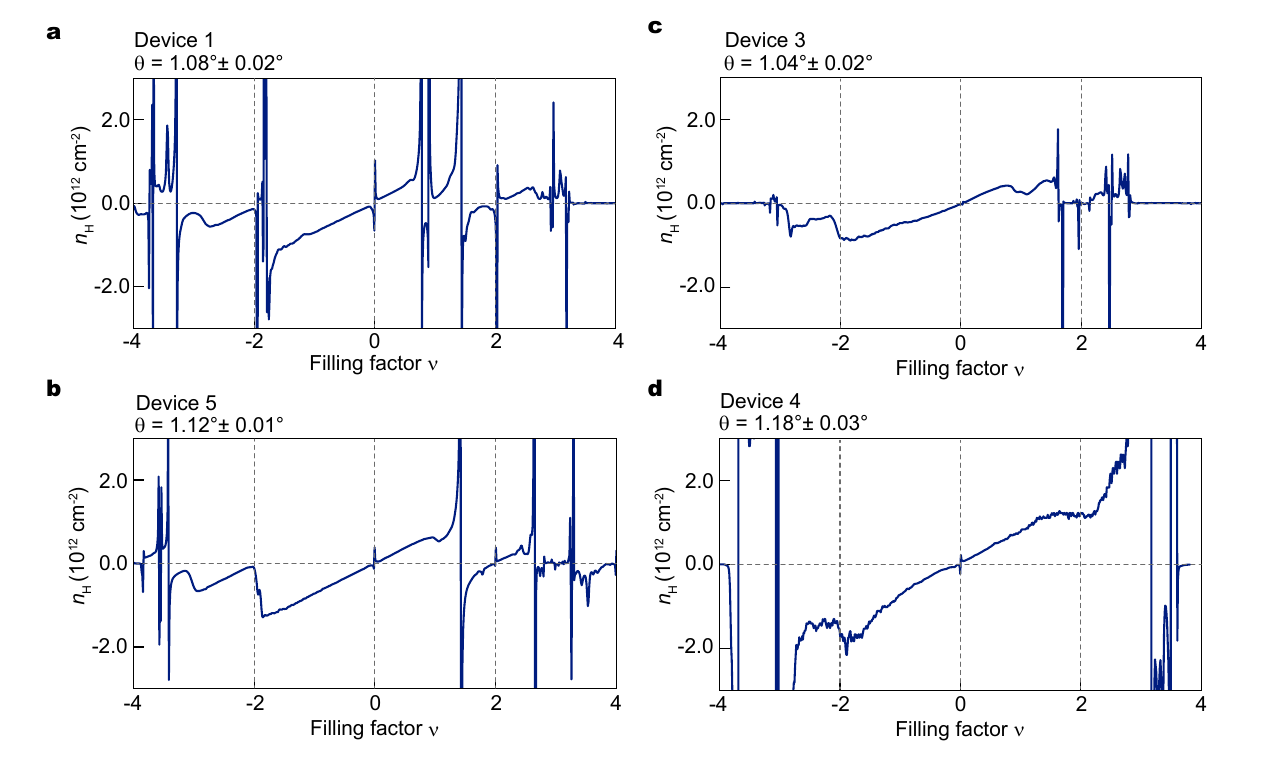}
\caption{\textbf{Hall density ($n_\mathrm{H}$) as a function of filling factor $\nu$.} The Hall effect measurements are performed at 0.8 K and 0.5 T. The vertical dashed lines show $\nu = 0, -2$ and $2$ filling and horizontal dashed line show the zero density. Device geometry precluded Hall measurements for Device 2.}
\label{fig:Hall}
\end{figure*}

\begin{figure*}[ht!]
\includegraphics[width= 7.2in]{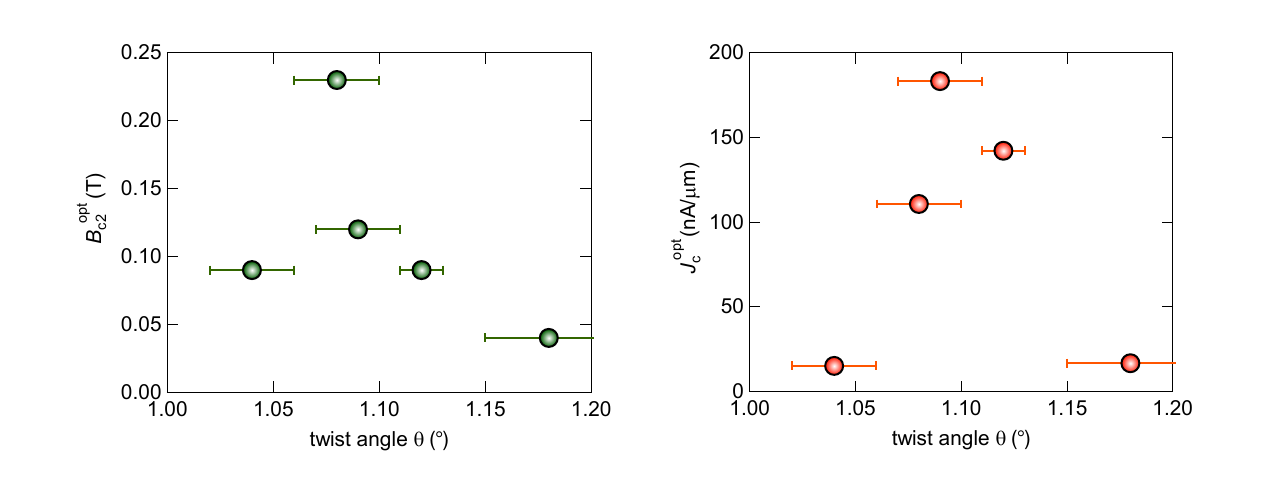}
\caption{\textbf{Upper critical field $B_\mathrm{c2}^\mathrm{opt}$ and critical current $J_\mathrm{c}^\mathrm{opt}$ at optimal doping as a function of twist angle.} $J_\mathrm{c}^\mathrm{opt}$ is a normalized value of the measured critical current by the channel width of each device. These values are extracted from the data measured at 10 mK.}
\label{fig:Bc2Icangle}
\end{figure*}

\end{document}